\documentclass[twocolumn]{aastex631}

\usepackage{amsmath}
\usepackage{subfigure}

\shortauthors{Xue et al.}

\begin{document}

\title{On the origin of a possible hard VHE spectrum from M87 discovered by LHAASO}

\author[0000-0003-1721-151X]{Rui Xue}
\altaffiliation{Corresponding authors}
\affiliation{Department of Physics, Zhejiang Normal University, Jinhua 321004, China; \textcolor{blue}{ruixue@zjnu.edu.cn}}

\author{Jia-Chun He}
\affiliation{Guangxi Key Laboratory for Relativistic Astrophysics, School of Physical Science and Technology, Guangxi University, Nanning 530004, China;}

\author[0000-0002-6809-9575]{Dingrong Xiong}
\altaffiliation{Corresponding authors}
\affiliation{Yunnan Observatories, Chinese Academy of Sciences, 396 Yangfangwang, Guandu District, Kunming, 650216, People's Republic of China; \textcolor{blue}{xiongdingrong@ynao.ac.cn}}%
\affiliation{Center for Astronomical Mega-Science, Chinese Academy of Sciences, 20A Datun Road, Chaoyang District, Beijing, 100012, People's Republic of China}
\affiliation{Key Laboratory for the Structure and Evolution of Celestial Objects, Chinese Academy of Sciences, 396 Yangfangwang, Guandu District, Kunming, 650216, People's Republic of China}

\author[0000-0002-3883-6669]{Ze-Rui Wang}
\altaffiliation{Corresponding authors}
\affiliation{College of Physics and Electronic Engineering, Qilu Normal University, Jinan 250200, China; \textcolor{blue}{zerui\_wang62@163.com}}%


\begin{abstract}
Recent LHAASO observations hint at potential spectral hardening around 20 TeV in M87's very high energy (VHE) emission, suggesting a possible new radiation component. In this work, we construct averaged multiwavelength SEDs by combining data from Chandra and Swift-UVOT/XRT covering the same period as the LHAASO detection to investigate the origin of this feature. We test several radiation mechanisms, including the $pp$ interaction, proton synchrotron emission, photomeson process and two-zone leptonic model. We find that only the pion decay gamma rays in $pp$ interactions can interpret this feature in the framework of the one-zone model. With analytical analysis, we prove that proton synchrotron emission cannot generate a hard spectrum above 0.17~TeV. For photomeson model, it requires an emission zone compressed near the Schwarzschild radius of the central supermassive black hole, incompatible with broadband optical-GeV spectral constraints. In addition, the two-zone leptonic model also emerges as a viable alternative. 
\end{abstract}


\section{Introduction}
M87, the central giant elliptical galaxy of the Virgo Cluster at a distance of 16.4 Mpc \citep{2010A&A...524A..71B}, is one of the nearest active galactic nuclei (AGN). As a Fanaroff--Riley (FR) I radio galaxy, it hosts a supermassive black hole (SMBH) with a mass of $\sim6.5\times10^9~M_\odot$ whose shadow was first directly imaged by the Event Horizon Telescope \citep{2019ApJ...875L...6E}. Its prominent relativistic jet, extending beyond the kpc scale with a $15^\circ-25^\circ$ viewing angle and displaying superluminal motion and complex knot structures, has been studied across electromagnetic wavelengths for over a century \citep[see][for a review]{2024A&ARv..32....5H}. M87 holds particular significance as the first radio galaxy detected at very high energies \citep[VHE, $>100\rm~GeV$;][]{2003A&A...403L...1A}, with flaring activity observed in 2005 \citep{2006Sci...314.1424A}, 2008 \citep{2008ApJ...685L..23A, 2009Sci...325..444A}, 2010 \citep{2012ApJ...746..151A}, 2018 \citep{2024A&A...685A..96H} and 2022 \citep{2024ApJ...975L..44C} showing day-scale variability, which is a key constraint implying compact VHE emission regions.

As a typical FR I radio galaxy exhibiting no broad emission lines, M87's broadband spectral energy distribution (SED) is usually interpreted by the one-zone synchrotron self-Compton (SSC) model \citep[e.g.,][]{2009ApJ...707...55A}. Due to the large ratio between the high-energy and low-energy peak frequencies, i.e., the significant separation between these peaks, the magnetic field introduced in the one-zone SSC model is relatively weak, usually on the order of a milligauss \citep{2009ApJ...707...55A, 2015MNRAS.450.4333D}. Even so, due to the spectral softening imposed by the Klein-Nishina (KN) effect, some studies suggest that SSC radiation may not be able to interpret the VHE spectrum \citep{2005ApJ...634L..33G, 2016ApJ...830...81F, 2020MNRAS.492.5354M}. Moreover, a possible excess over the standard power-law model above $\sim10\rm~GeV$ during the low-state challenges the one-zone SSC model and suggests the existence of a second radiation component \citep{2019A&A...623A...2A}. To better account for the VHE radiation observed in M87, both multi-zone leptonic models \citep{2005ApJ...634L..33G, 2008A&A...478..111L, 2008MNRAS.385L..98T} and hadronic processes \citep{2016ApJ...830...81F, 2022ApJ...934..158A, 2022PhRvD.106j3021X} have been proposed.

Very recently, the Large High Altitude Air Shower Observatory (LHAASO) collaboration find that the VHE spectrum of M87 extends to 20 TeV, accompanied by a tentative spectral hardening around $\sim20\rm~TeV$, while this hardening feature remains statistically marginal \citep{2024ApJ...975L..44C}. This kind of hard VHE spectrum has previously been exclusively detected in jetted AGNs, i.e., BL Lacs, with relativistic jets oriented toward Earth. Consequently, if the hard VHE spectrum of M87 is genuine, this feature would necessitate alternative radiation mechanisms beyond the canonical SSC paradigm, potentially indicative of hadronic processes or complex multi-zone scenarios. In this work, to understand the radiation mechanism of M87 comprehensively, we build averaged multiwavelength SED by combining
observations of Chandra, $Swift$-XRT and $Swift$-UVOT. Several models are tested to interpret the SED, especially the hard VHE spectrum. In Section~\ref{data}, we present the detailed $Swift$-XRT and $Swift$-UVOT data reduction. The modeling results are shown in Section~\ref{result}. Finally, we end with discussion and conclusion in Section~\ref{DC}. Throughout the paper, the cosmological parameters $H_{0}=69.6\ \rm km\ s^{-1}Mpc^{-1}$, $\Omega_{0}=0.29$, and $\Omega_{\Lambda}$= 0.71 are adopted \citep{2014ApJ...794..135B}.

\section{Data Analysis}\label{data}
In this section, we present the multiwavelength observations of M87 from 2021 March 5 to 2024 March 16 during the operation of the Water Cherenkov Detector Array (WCDA) of LHAASO. The spectra and light curves of Fermi-LAT and LHAASO-WCDA are given in \cite{2024ApJ...975L..44C}. To derive the broadband light curves and averaged multiwavelength SED, the data reduction processes of Chandra, Swift-XRT, and Swift-UVOT are described in the following.

\subsection{Chandra}
The Chandra X-ray Observatory, launched in 1999, provides X-ray imaging and spectroscopy with high angular resolution ($< 0.5''$) in the energy range 0.1--10 keV \citep{2002PASP..114....1W}. Its Science Instrument Module contains two focal plane instruments, the Advanced CCD Imaging Spectrometer (ACIS) and the High Resolution Camera (HRC). The ACIS module is used for spectral analysis. In this paper, the extraction of spectrum and light curve is performed using CIAO (v4.16) and the Chandra Calibration Database (CALDB v4.11.2).

We analyse the Chandra-ACIS data following the guidance of Science Threads\footnote{\url{https://cxc.harvard.edu/ciao/threads/index.html}}. We select the observational informations of M87 cover from March 8, 2021 (ObsID 23669), to March 16, 2024 (ObsID 25369), the total exposure time is over 66.24 kiloseconds. To reduce uncertainties caused by the position offsets of different observations, we perform the astrometric corrections. The count images, exposure maps and weighted PSF maps are produced by the $fluximage$ and $mkpsfmap$ tools. We used the $wavdetect$ tool to obtain the position of the target source. We selected an observation with the longest exposure time as a reference and performed cross-matching between this reference and the other observations; we used $wcs\_match$ to generate the transformation matrix and the $wcs\_update$ tool to update the coordinates of the shorter observations.

For spectral analysis, the specextract tool is used to perform an accurate aperture photometry of an elliptical nucleus of $0.70''\times0.58''$ \citep{2005ApJ...627..140P}. The background region is defined as $25''\times3''$ rectangular regions situated $5''$ north and south of the jet. We use the sherpa\footnote{\url{https://cxc.harvard.edu/sherpa/threads/index.html}} package to simultaneously fit the broad-band spectra (0.5--7 keV) of multi-observations with a single power law plus a Galactic absorption model and pile-up model $jdpileup$ \citep{2001ApJ...562..575D}. The absorption column density $N_H$ is set to free. The errors of flux and photon index are evaluated at 90 \% confidence level. In Fig.~\ref{SED}, we show the X-ray light curve and spectrum of the nuclear region, with a spectral index of $1.38\pm0.13~(F_\nu\propto \nu^{-\alpha})$.

\subsection{\textit{Swift-XRT}}
The average spectrum in photon-counting (PC) mode, combined from multiple observations, is derived from Swift X-Ray Telescope (\textit{XRT}) data products generator\footnote{\url{https://www.swift.ac.uk/user\_objects/}} (\textit{xrt\_prods}), which corrects for pile up, CCD bad columns, and other instrumental artifacts \citep{Evans2009}. For spectral extraction, a circular source region with a diameter of $35''$ was selected, following the methodology of \cite{Algaba2024}. \textit{Swift-XRT} cannot spatially resolve the inner jet, including the core and HST-1 knot; therefore, the background-subtracted fluxes should be considered as upper limits for the X-ray emission from the core, the HST-1 knot, and the jet \cite[see][]{Algaba2024}. We apply the \textit{grppha} command rebin channels, with a minimum number of groups greater than 25. The spectra are fitted using\textit{ XSPEC} (\textit{version 12.9}) with the model \textit{TBabs*(vvapec + cflux*(po + po + po))}, where the three power-law (\textit{po}) components represent the core, the HST-1 knot, and the jet emission, respectively, and vvapec accounts for background emission. During the fitting process, the \textit{po} index is fixed at 2.0, following the approach of \cite{Algaba2024}. The chi-squared value over degrees of freedom (\textit{dof}) for the fit is 178/163. The non-absorbed flux (\textit{cflux}) in the 2 to 10 keV range, with an approximate median of 6 keV, is used as the upper limit.

\subsection{$Swift-UVOT$}
Swift Ultraviolet and Optical Telescope ($UVOT$) provides ultraviolet ($UVW1$, $UVM2$, $UVW2$) and optical ($V$, $B$, $U$) data \citep{Roming2005}, which are retrieved from the NASA High-Energy
Astrophysics Archive Research Center (HEASARC\footnote{\url{https://heasarc.gsfc.nasa.gov/cgi-bin/W3Browse/swift.pl}}) within the time interval specified in the section 2.1. The observational data are collected between March 19, 2021, and November 22, 2023, comprising a total of 62 observations. Of these, 3 are discarded due to more than $10'$ positional offsets that exclude M87 from the field of view. Data processing is performed using \textit{HEASoft} version 6.32.1 and calibration files (\textit{CALDB}) version 20211108. 

We verify the alignment of level 2 images to the World Coordinate System. The software automatically conducts a small-scale sensitivity check. The \textit{uvotimsum} command is used to combine image extensions, while \textit{uvotsource} performs aperture photometry using a circular source region with a $5''$ radius and a circular background region with an radius of $30''$, following the setup outlined in \cite{Algaba2021}. The flux density is corrected for Galactic extinction following the method described by \cite{2024ApJS..271...10W}. The SEDs spanning the optical to ultraviolet energy range are constructed using the average flux density and average errors over the selected period.

\begin{figure}[htbp]
\subfigure{
\includegraphics[width=0.55\textwidth]{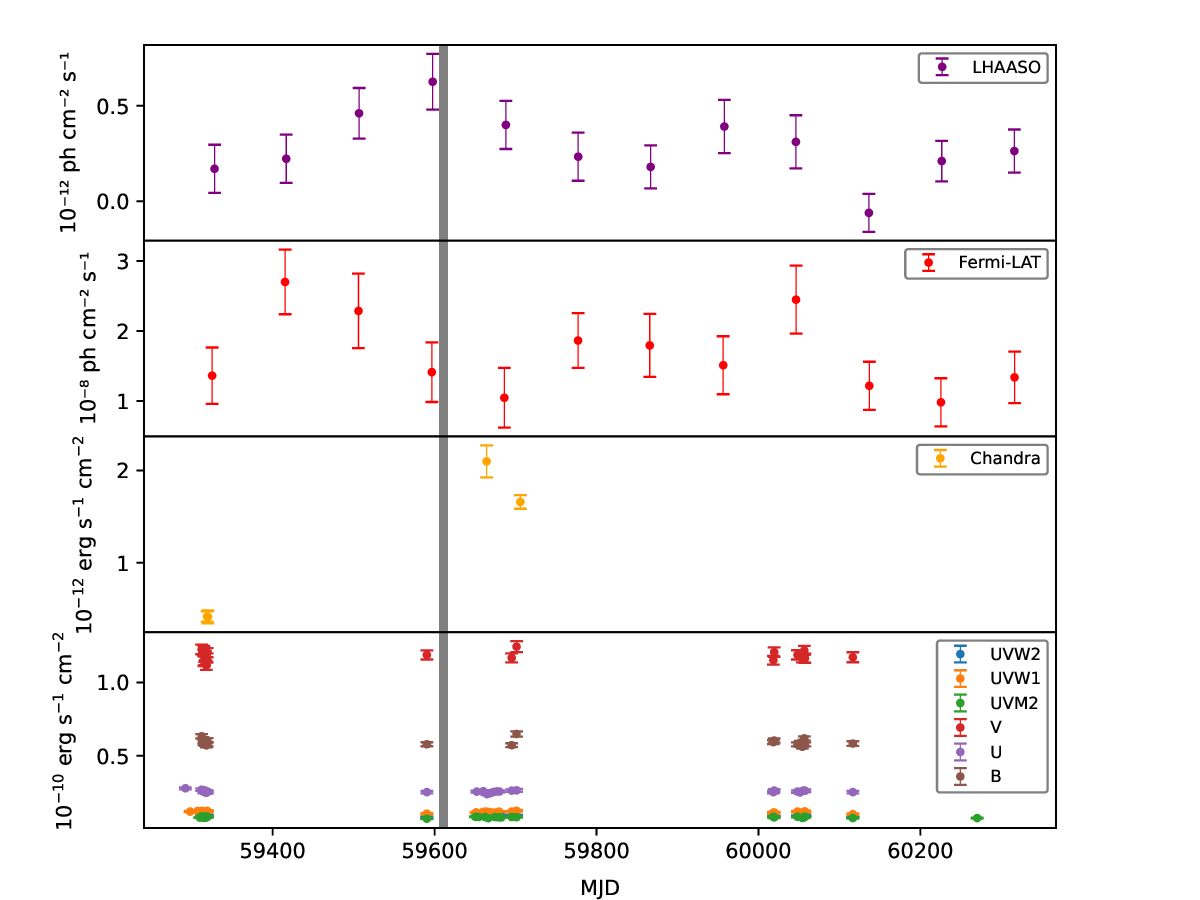}
}\hspace{-5mm}
\quad
\subfigure{
\includegraphics[width=0.5\textwidth]{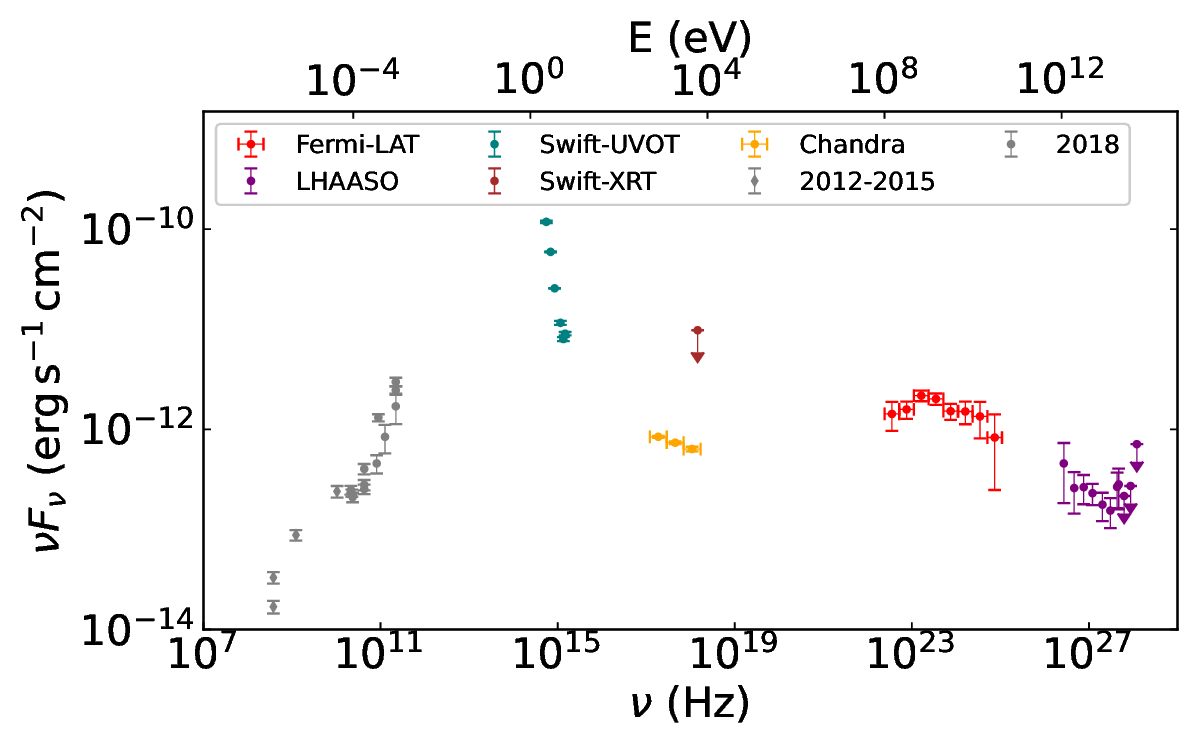}
}
\caption{Upper panel: Multiwavelength light curves (LCs) of M87. Panels from top to bottom: LCs of LHAASO, Fermi-LAT, Chandra, and Swift-UVOT. The LHAASO-WCDA flare period from MJD 59607 to MJD 59615 is shown as the gray vertical shaded region. Lower panel: The average broadband SED of M87 from 2021 March 8 to 2024 March 16. The meaning of all data points is explained in the inset legends. Please note that the radio data are historical observations, with the 2018 measurements obtained from \cite{Algaba2024} and the 2012-2015 data taken from \cite{2020MNRAS.492.5354M}.
\label{SED}}
\end{figure}

\section{Modeling Results}\label{result}
As shown in the upper panel of Fig.~\ref{SED}, M87 does not show a significant flare during the same observation period of LHAASO in all bands, except for the GeV and TeV flares from MJD 59607 to MJD 59615 as reported by \cite{2024ApJ...975L..44C}. The gray vertical shaded region in this panel, which denotes these GeV and TeV flares, confirms the absence of simultaneous multiwavelength observations during this period, consistent with the findings of \cite{2024ApJ...975L..44C}. Here, we focus on studying the origin of the potential spectral hardening around 20 TeV detected by LHAASO during its long-term average state and low state (i.e., excluding the flare period from MJD 59607 to MJD 59615). Given that the low state differs from the long-term average state only by the exclusion of this brief seven day interval, their resulting SEDs are basically the same. Therefore, here we use the long-term averaged flux of each band to construct the SED. Furthermore, to ensure that the model-predicted radio flux does not overshoot the radio observation, historical broadband radio data (from $\sim 10^8~\rm Hz$ to $\sim 10^{11}~\rm Hz$) given by \cite{2020MNRAS.492.5354M} and \cite{Algaba2024} are also included in the SED. The multiwavelength SED of M87 is shown in the lower panel of Fig.~\ref{SED}. It can be seen that the SED of M87 shows remarkable discrepancies between the optical/UV bands and other wavebands. This suggests a dominant contribution of the optical/UV bands from stellar populations inherent to the host galaxy rather than from AGN-related processes. We therefore use the SWIRE template\footnote{\url{http://www.iasf-milano.inaf.it/~polletta/templates/swire_templates.html}} \citep{2007ApJ...663...81P} for elliptical galaxies, corresponding to an age of 13 Gyr, to reproduce the optical/UV spectrum.

\begin{table*}
\setlength\tabcolsep{2.8pt}
\caption{Parameters for SED Fitting with the One-zone SSC+$pp$ Model.}
\centering
\begin{tabular}{cccccccccccc}
\hline\hline
Free parameters		&	$R$~(cm)	&	$B$~(G)	&	$L_{\rm e,inj}~(\rm erg~s^{-1})$	&	$\gamma_{\rm e,b}$	&	$p_{\rm e,1}$	&	$p_{\rm e,2}$	&	$\xi$	&	$\gamma_{\rm p,max}$	&	$p_{\rm p}$	\\
\hline
Values 		&	$1.3\times10^{18}$	&	$9.0\times10^{-4}$	&	$1.2\times10^{43}$	&	$4.0\times10^4$	&	1.9	&	3.4	&	$3.5\times10^{-3}$	&	$6.2\times10^5$	&	2.0	\\
\hline\hline
Derived/Fixed parameters	&	$\Gamma$ &	$\theta_{\rm obs}~(\circ)$	&	$\delta_{\rm D}$	&	$\gamma_{\rm e,min}$	&	$\gamma_{\rm e,max}$	&	$\gamma_{\rm p,min}$	&	$\chi_{\rm p}$	&	$n_{\rm H}~(\rm cm^{-3})$	&	$f_{\rm pp}$	&	\\
\hline																							
Values	&	3 &	15	&	3.7	&	1	&	$10^8$	&	1	&	0.5&	$4.5\times10^{2}$	&	$8.4\times10^{-6}$	&	\\
\hline
\label{ppparameters}
\end{tabular}
\textbf{Notes.} The blob radius $R$ is not a completely free parameter. In order to make the emission from $pp$ interactions contribute to VHE bands, the adopted values of $R$ are lower than the maximum value constrained by Eq.~(\ref{RS}). Values of $\Gamma$ and $\theta_{\rm obs}$ are fixed following the apparent motion observations. For simplicity, we fix $\gamma_{\rm e,min}=1$, $\gamma_{\rm e,max}=10^8$, and $\gamma_{\rm p,min}=1$, since they have minor impacts on our modeling result. As suggested in \cite{2022A&A...659A.184L}, we fix $\chi_{\rm p}=0.5$. It means that the maximum parameter space can be obtained under this condition, which also indicates that the kinetic power of cold protons and the relativistic proton injection power each account for half of the total jet power. The number density of cold protons $n_{\rm H}$ is derived by $n_{\rm H}=\frac{(1-\chi_{\rm p})\xi L_{\rm Edd}}{\pi R^2\Gamma^2m_{\rm p}c^3}$, where $\xi$ is the ratio of the total jet power to $L_{\rm Edd}$, and the $pp$ interaction efficiency $f_{\rm pp}$ is obtained by setting a constant cross section, i.e., $\sigma_{\rm pp}=6\times10^{-26}~\rm cm^2$, for inelastic $pp$ interactions. $\gamma_{\rm p,max}$ is not entirely a free parameter. Under the combined influence of $B$ and $R$, we have $\gamma_{\rm p,max}\leqslant5.6\times10^7(\frac{B}{10^{-3}\rm~G})(\frac{R}{10^{18}\rm~cm})\approx6.6\times10^7$. As long as the value of $\gamma_{\rm p,max}$ is chosen within the range of $5.8\times10^4$ and $6.6\times10^7$, the modeling result remains physically self-consistent.
\end{table*}

The hardening observed in M87's low-state VHE spectrum by LHAASO at $\sim 20\rm~TeV$ directly indicates the presence of additional radiation components. Here, we explore potential radiation mechanisms extending beyond the canonical one-zone SSC paradigm. For all subsequent models to be tested, it is uniformly assumed that non-thermal radiation originates from the spherical emission region with radius $R$, which is filled with a uniform entangled magnetic field $B$ and a plasma of charged particles. The emitting region moves along a relativistic jet with the bulk Lorentz factor $\Gamma=(1-\beta^2)^{-1/2}$ at a viewing angle $\theta \mathrm{^{obs} }$ to the observer's line of sight, where $\beta c$ is the speed of the emitting region. In the modeling, we fix $\theta^{\rm obs} = 15^\circ$ and $\Gamma=3$, which are consistent with apparent motion observations of M87 \citep{1999ApJ...520..621B, 2007ApJ...660..200L, 2013ApJ...774L..21M}. The Doppler beaming factor $\delta_{\rm D}$ can be obtained through $\delta_{\rm D} =\left [ \Gamma \left ( 1-\beta \cos \theta ^{\mathrm{obs}  }  \right )  \right ] ^{-1}\approx 3.7$. It should be noted that M87's jet has been observed to acceleration along the jet direction toward large scales \citep[e.g.,][]{2016A&A...595A..54M, 2019ApJ...887..147P}. In the Appendix~\ref{APP}, we present supplementary fitting results with a larger value of $\Gamma$. In this section, parameters without superscript are measured in the comoving frame, and those with superscript `obs' are measured in the frame of the observer, unless specified otherwise.

\subsection{One-Zone SSC+$pp$ Model}\label{SSCpp}
The $pp$ interaction in the M87 jet are rarely considered, as efficient $pp$ interactions would require a substantial kinetic power of cold protons in the jet, potentially exceeding the Eddington luminosity of the SMBH. \cite{2012ApJ...755..170B} proposed that a star captured by and colliding with the jet could supply dense cold matters, and that the gamma rays produced via $pp$ interactions could thereby explain the rapid gamma-ray flare of M87. 
It should be noted that this jet-star collision model is more suitable for explaining fast variabilities. Since LHAASO detected the hard TeV spectrum in the average/low state, it is more plausible that the jet itself provides the target particles. \cite{2011A&A...531A..30R} proposed a conical jet model and demonstrated that $pp$ interactions can be efficient in sub-Eddington jets, since protons with long cooling timescales can continuously produce radiative output while propagating along the jet. Their fitting result of M87 reveals that gamma rays from $\pi^0$ decay dominate the GeV-TeV spectrum and manifest a soft spectrum in the TeV band. For the one-zone model, our previous study \citep{2022PhRvD.106j3021X} proved that $pp$ interactions can be significant in the sub-Eddington jet. Our modeling result predicted that a hard VHE spectrum from $\pi^0$ decay could be detected by LHAASO.

For the one-zone model, we first constrain the physical parameters using an analytic method. Following our previous study \citep{2022PhRvD.106j3021X}, if $\pi^0$ decay gamma rays in $pp$ interactions has contribution to the LHAASO detected VHE spectrum, and the introduced jet power does not exceed the Eddington luminosity of the SMBH, the blob radius will be strictly constrained by
\begin{equation}\label{RS}
R\leqslant R_{\rm S}\frac{\sigma_{\rm pp}}{12\sigma_{\rm T}}\frac{\delta_{\rm D}^4}{\Gamma^4}\frac{L_{\rm Edd}}{L_{\rm TeV}^{\rm obs}}\approx 4\times10^{21}\rm~cm,
\end{equation}
where $\sigma_{\rm pp}=6\times10^{-26}~\rm cm^2$ is the cross section for inelastic $pp$ interactions, $\sigma_{\rm T}$ is the Thomson scattering cross section, $R_{\rm S}\approx2\times10^{15}~\rm cm$ is the Schwarzschild radius of the SMBH, $L_{\rm Edd}\approx9\times10^{47}\rm~erg~s^{-1}$ is the Edington luminosity of the SMBH, and $L_{\rm LHAASO}^{\rm obs}\approx 8.1\times10^{39}\rm~erg~s^{-1}$ is the hard VHE luminosity. It can be seen that a substantial parameter space exists, allowing the one-zone SSC+$pp$ model to account for the broadband SED of M87 reasonablely.

\begin{figure}[htbp]
\includegraphics[width=0.5\textwidth]{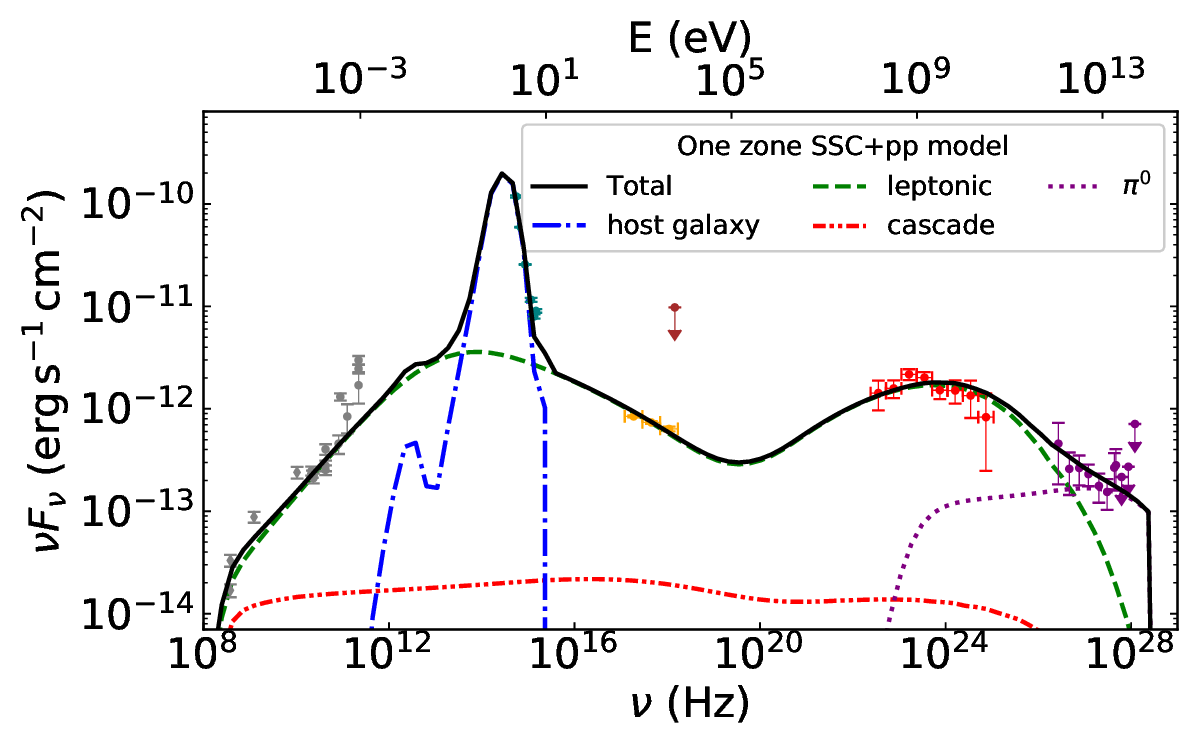}
\caption{One-zone SSC+$pp$ modeling. The dashed green curve represents the leptonic emission, including synchrotron and SSC emission, from primary relativistic electrons. The dotted purple curve represents the gamma-ray emission from $\pi^0$ decay in $pp$ interactions. The dot-dashed red curve represents the leptonic emission from pair cascades. The solid black curve is the total emission from the blob.
\label{pp}}
\end{figure}

In the modeling, primary relativistic electrons are assumed to be injected with a smooth broken power-law energy distribution at a constant rate given by $\dot{Q}^{\rm inj}_{\rm e}(\gamma_{\rm e})=\dot{Q}_{\rm e,0}\gamma_{\rm e}^{-p_{\rm e,1}}[1+(\frac{\gamma_{\rm e}}{\gamma_{\rm e,b}})^{(p_{\rm e,2}-p_{\rm e,1})}]^{-1}H(\gamma_{\rm e};\gamma_{\rm e,min},\gamma_{\rm e,max})$, where $H(x;x_1,x_2)$ for $x_1\leq x\leq x_2$, $\gamma_{\rm e,min/b/max}$ represent the minimum, break, and maximum Lorentz factors, and $p_{\rm e,1/2}$ represent the spectral indices before and after $\gamma_{\rm e,b}$. Given an electron injection luminosity $L_{\rm e,inj}$, $\dot{Q}_{\rm e,0}$ can be determined by $\int \gamma_{\rm e}m_{\rm e}c^2 \dot{Q}^{\rm inj}_{\rm e}(\gamma_{\rm e}) d\gamma_{\rm e}=3L_{\rm e,inj}/4\pi R^3$, where $m_{\rm e}$ is the electron rest mass. When the injection of electrons is balanced with radiative cooling and/or particle escape, a steady-state electron energy distribution is achieved. Synchrotron, and SSC emissions are considered in the radiative cooling. For relativistic protons, they are assumed to be injected with a power-law energy distribution at a constant rate given by $ \dot{Q}^{\rm inj}_{\rm p}(\gamma_{\rm p})=\dot{Q}_{\rm p,0}\gamma_{\rm p}^{-p_{\rm p}}H(\gamma_{\rm p};\gamma_{\rm p,min},\gamma_{\rm p,max})$, where $\gamma_{\rm p,min/max}$ are the minimum, and maximum proton Lorentz factors, $p_{\rm p}$ is the spectral index. Similarly, by giving a proton injection luminosity $L_{\rm p,inj}$, $\dot{Q}_{\rm p,0}$ can be obtained. By taking into account the cooling of $pp$ interactions, the proton-synchrotron cooling and the escape of the protons, the steady-state proton density distribution can be derived. When applying the one-zone SSC+$pp$ model, it is expected that the SED below TeV energies to be dominated by leptonic emission from primary electrons, while the VHE emission is attributed to gamma rays from $\pi^0$ decay in $pp$ interactions. The $\pi^0$ decay gamma rays take about 10\% of the energy of parent protons, therefore it is required that $\gamma_{\rm p,max}\geqslant5.8\times10^4(\frac{20\rm~TeV\times10}{200\rm~TeV})(\frac{3.7}{\delta_{\rm D}})$. It is necessary to check whether protons can be accelerated to energies high enough to account for the hard spectrum at $\sim20\rm~TeV$ detected by LHAASO. In our modeling, we do not explicitly specify the particle acceleration mechanism. Instead, we assume that relativistic protons, already accelerated, are injected into the emission region. Here we may assume that the dominant acceleration mechanism is the diffusive shock acceleration, then the acceleration timescale $t_{\rm acc}$ can be evaluated by $t_{\rm acc}\simeq \frac{20\alpha \gamma_{\rm p}m_{\rm p}c}{3eB}$, where $\alpha \geqslant1$ is the parameter which depends on the spectrum of magnetic turbulence and on the velocity of the upstream-flow \citep[e.g.,][]{2007Ap&SS.309..119R}. As shown by \cite{2022PhRvD.106j3021X}, the $pp$ cooling timescale within the jet typically far exceeds the escape timescale $t_{\rm esc}=R/c$, implying the interaction efficiency $f_{pp}\ll1$. Consequently, $\gamma_{\rm p,max}$ can be obtained by equating $t_{\rm acc}$ and $t_{\rm esc}$, and its upper limit is reached under the default assumption $\alpha = 1$, i.e., 
\begin{equation}\label{max}
\gamma_{\rm p,max}\leqslant5.6\times10^7(\frac{B}{10^{-3}\rm~G})(\frac{R}{10^{18}\rm~cm}). 
\end{equation}
Therefore, when setting $\gamma_{\rm p,max}$ in the following modeling, we ensure that its value does not exceed the upper limit determined for the given values of $B$ and $R$.

After obtaining the steady-state primary electron and proton energy distributions, we apply analytical expressions developed by \cite{2006PhRvD..74c4018K} to obtain the differential spectrum of decayed gamma rays and secondary electrons/positrons (pairs), evaluate the gamma rays induced pair cascade using the method developed in \cite{1983Afz....19..323A}, and calculate the synchrotron and SSC emissions from primary relativistic electrons and secondary pairs using the public \texttt{NAIMA} Python package \citep{naima}. A more detailed numerical model description can be found in our previous work \citep{2022PhRvD.106j3021X}. Table~\ref{ppparameters} summarizes the model parameters, and the broadband SED is shown in Fig.~\ref{pp}. It can be seen that the broadband spectrum from optical to GeV band is contributed by the leptonic emission from the primary electrons (green dashed curves) and hard VHE spectrum is dominated by the $\pi^0$ decay gamma rays from $pp$ interactions (purple dotted curves). Therefore, we propose that one-zone SSC+$pp$ model provides a reasonable explanation for the broadband SED of M87, particularly accounting for the hard VHE spectrum observed by LHAASO.

\subsection{Proton Synchrotron Model}
\cite{2003APh....19..559P} initially predicted that M87 might produce detectable VHE radiation through proton synchrotron (p-syn) radiation and induced cascade emission. Subsequently, \cite{2004A&A...419...89R} conducted detailed modeling and found that p-syn radiation under varying magnetic field strengths, along with related cascade emission, could contribute significantly to VHE emissions. However, limited by sparse observational data, their modeling results failed to effectively account for emissions in the X-ray and GeV bands. Recently, \cite{2020MNRAS.492.5354M} also employed p-syn radiation to explain the high-energy bump of M87, successfully fitting the VHE spectrum. 

\begin{figure}[htbp]
\subfigure{
\includegraphics[width=0.5\textwidth]{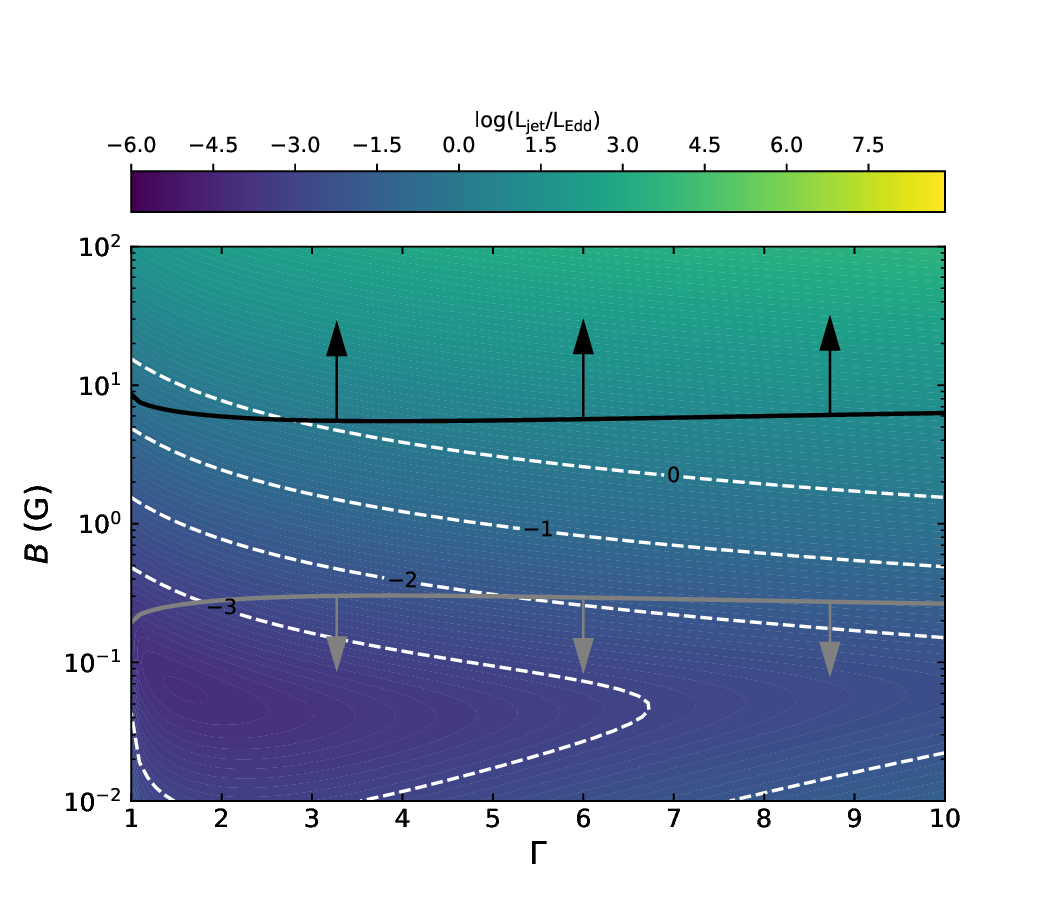}
}
\quad
\subfigure{
\includegraphics[width=0.5\textwidth]{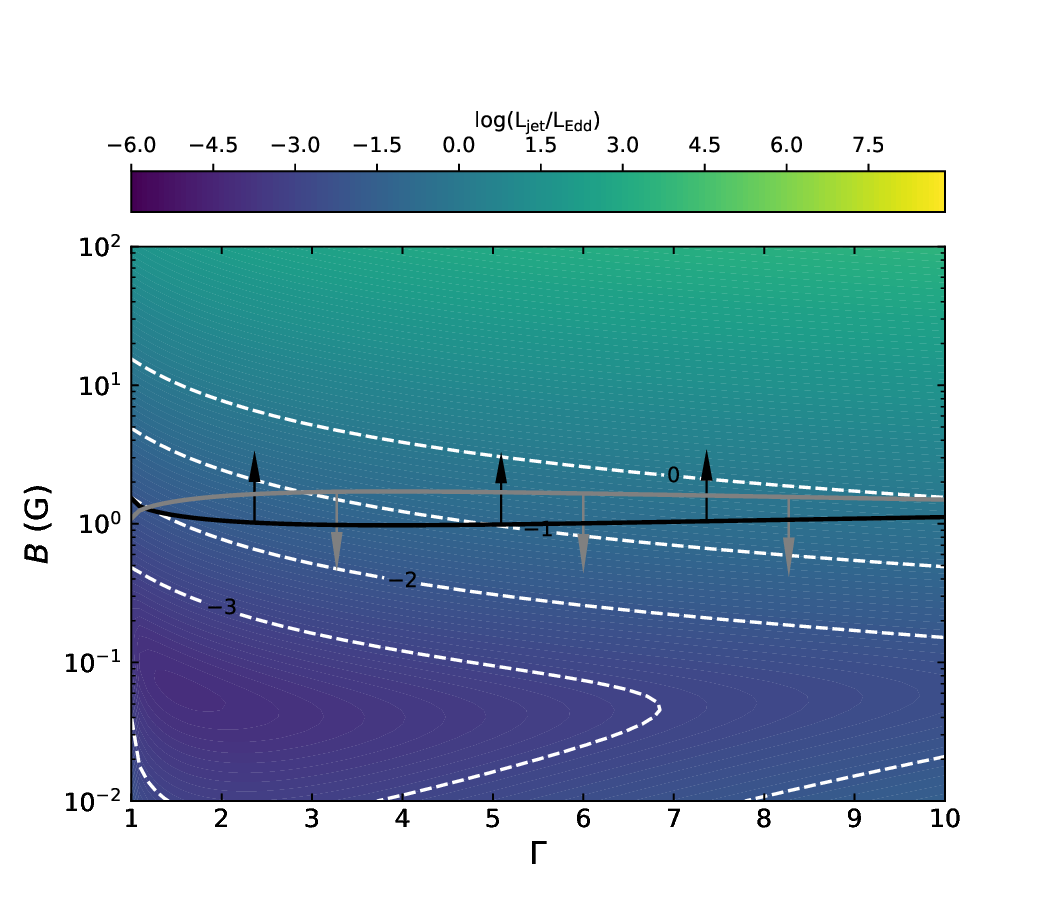}
}
\caption{The ratio of $L_{\rm jet}/L_{\rm Edd}$ in the $\Gamma-B$ diagram for the one-zone proton-synchrotron model in the case of $E_{\rm VHE}=20\rm~TeV$ (upper panel) and $E_{\rm VHE}=0.1\rm~TeV$ (lower panel). The black curves with arrows represent the parameter space that satisfied the Eq.~(\ref{BL}). The gray curves with arrows represent the parameter space that satisfied Eq.~(\ref{BU}). The white dashed contours denote specific values of log($L_{\rm jet}/L_{\rm Edd}$) associated with the color bar.
\label{psyn}}
\end{figure}

Our previous work \citep{2023PhRvD.107j3019X} proposed an analytical method to constrain the parameter space in the context of one-zone proton-synchrotron model. In the analytical method, the parameter space is constrained by requirement that the total jet power, dominated by the injection power of relativistic protons and the power carried in magnetic field, stays below the Eddington luminosity, by Hillas condition ensuring that relativistic protons can be accelerated to the required maximum energy \citep{1984ARA&A..22..425H}, and by the condition that the blob remains transparent to gamma rays. It is noteworthy that the maximum proton energy inferred from the Hillas condition implicitly assumes that particle escape dominates over radiative cooling. \cite{2024ApJ...975L..44C} has extended the VHE spectrum of M87 to approximately 20 TeV, higher than previously observed, which indicates that p-syn cooling may be more important than escape. Consequently, the analytical method of \cite{2023PhRvD.107j3019X} should be refined, and one must investigate whether protons can be accelerated to energies sufficient to account for a $\sim$20 TeV in radiative cooling-dominated or escape-dominated regimes. Let us first study the escape-dominated scenario. Assuming that the dominant acceleration mechanism is the diffusive shock acceleration, then we can obtain the maximum $\gamma_{\rm p, max}$ with Eq.~(\ref{max}). Since $\sim20\rm~TeV$ emission is expected from p-syn emission, the proton Lorentz factor $\gamma_{\rm p, VHE}$ can be estimated using the monochromatic approximation
\begin{equation}\label{gammaVHE}
\gamma_{\rm p, VHE}\approx 9\times10^{11}\big(\frac{E_{\rm VHE}}{20\rm~TeV} \big)^{1/2} \big(\frac{1\rm~G}{B} \big)\big(\frac{3.7}{\delta_{\rm D}} \big)^{1/2},
\end{equation}
where $E_{\rm VHE}$ represents the maximum photon energy emitted by p-syn radiation. To ensure that the observed VHE emission can be explained by p-syn radiation, $\gamma_{\rm p, VHE}$ should be lower than $\gamma_{\rm p, max}$. Then we have
\begin{equation}\label{BL}
B\geqslant 6\rm~G\big(\frac{E_{\rm VHE}}{20\rm~TeV} \big)^{1/3} \big(\frac{10^{18}\rm~cm}{R} \big)^{2/3}\big(\frac{3.7}{\delta_{\rm D}} \big)^{1/3}.
\end{equation}
On the other hand, the condition $t_{\rm esc}\leqslant t_{\rm cool}(\gamma_{\rm p, VHE})$ should be satisfied in the escape-dominated scenario, in which $t_{\rm cool}(\gamma_{\rm p, VHE})=\frac{6\pi m_{\rm e}c^2}{c\sigma_{\rm T}B^2\gamma_{\rm p, VHE}}(\frac{m_{\rm p}}{m_{\rm e}})^3$ is the synchrotron cooling timescale for protons with $\gamma_{\rm p, VHE}$. Then we have
\begin{equation}\label{BU}
B\leqslant 0.28\rm~G\big(\frac{E_{\rm VHE}}{20\rm~TeV} \big)^{-1/3} \big(\frac{10^{18}\rm~cm}{R} \big)^{2/3}\big(\frac{3.7}{\delta_{\rm D}} \big)^{-1/3}.
\end{equation}
By combining Eqs.~(\ref{BL}) and (\ref{BU}) with the analytical method of \cite{2023PhRvD.107j3019X}, we present the resulting parameter space in the upper panel of Fig.~\ref{psyn} by assuming $R=10^{18}\rm~cm$. It can be seen that, despite a broad parameter space satisfying the sub-Eddington jet power requirement, the constraints from Eqs.~(\ref{BL}) and (\ref{BU}) render no viable parameter space. Since the exponent of $R$ in both Eqs.~(\ref{BL}) and (\ref{BU}) is 2/3, adjusting $R$ cannot alter the conclusion that no viable parameter space exists. For the radiative cooling-dominated scenario, $t_{\rm acc}=t_{\rm cool}$ should be satisfied. Considering $\alpha=1$, we have
\begin{equation}\label{max1}
\gamma_{\rm p,max}\leqslant 8.3\times10^{10}(\frac{1\rm~G}{B}).
\end{equation}
Ensuring that the accelerated relativistic protons can explain the hard spectrum at $\sim20\rm~TeV$ requires that $\gamma_{\rm p,VHE} \leqslant \gamma_{\rm p,max}$. By substituting $E_{\rm VHE} = 20\rm~TeV$ and $\delta_{\rm D} = 3.7$ into Eq.~(\ref{gammaVHE}), one readily finds that the resulting $\gamma_{\rm p,VHE}$ is larger than $\gamma_{\rm p,max}$ by an order of magnitude. Therefore, p-syn radiation fails in the radiative cooling-dominated scenario either. For radio galaxies, $\delta_{\rm D}$ is confined to a very narrow range, and joint solutions of Eqs. (\ref{BL}) and (\ref{BU}), and of Eqs. (\ref{gammaVHE}) and (\ref{max1}) both show that p-syn emission is only viable for $E_{\rm VHE}\leqslant0.17\rm~TeV$. Consequently, p-syn emission cannot account for the spectral hardening around 20 TeV, while it can reproduce the soft VHE spectrum as in previous studies \citep[e.g.,][]{2020MNRAS.492.5354M}. For example, if assuming $E_{\rm VHE}=0.1\rm~TeV$, the derived parameter space is given in the lower panel of Fig.~\ref{psyn}. Overall, we conclude that p-syn radiation cannot account for the hard spectrum detected by LHAASO.

\subsection{Photomeson Model}
In this subsection, we test if $\pi^0$ decay gamma rays in the photomeson interactions could explain the hard VHE spectrum of M87 as suggested by some previous studies \citep{2015EPJC...75...52S, 2016ApJ...830...81F, 2022ApJ...934..158A}. 

If considering the peak cross section of photomeson interactions due to the $\bigtriangleup^+(1232)$ resonance, the energy of target photons $E_{\rm tar}^{\rm obs}$ in the observer' frame can be obtained through  
\begin{equation}
    E_{\rm tar}^{\rm obs} \simeq 21~\rm keV \big(\frac{\delta_D}{3.7} \big)^2 \big(\frac{20~\rm TeV}{E_{\rm VHE}^{\rm obs}}\big),
\end{equation}
where $E_{\rm VHE}^{\rm obs}\approx 20~\rm TeV$ is the energy of $\pi^0$ decay VHE photons, using the $\delta-$approximation. By taking $L_{\rm Edd}$ as the maximum proton injection luminosity, $10^{-28}~\rm cm^2$ as the photopion cross section weighted by inelasticity, the lower limit of flux of the target photons can be estimated by \citep{2024ApJS..271...10W}
\begin{equation}
\begin{split}
R\simeq 4\times10^{15}~\rm cm \big(\frac{{\nu F_{\nu}}_{\rm tar}^{\rm obs}(E_{\rm tar}^{\rm obs})}{4\times10^{-13}~\rm erg~s^{-1}~cm^{-2}} \big)\big(\frac{3}{\Gamma} \big)^2 \big(\frac{3.7}{\delta_{\rm D}} \big) \\ \big(\frac{E_{\rm VHE}^{\rm obs}}{20~\rm TeV} \big)
    \big(\frac{2\times10^{-13}~\rm erg~s^{-1}~cm^{-2}}{{\nu F_{\nu}}_{\rm 20~TeV}^{\rm obs}} \big)\big(\frac{M_{\rm BH}}{6.5\times10^9~M_{\odot}} \big),
\end{split}
\end{equation}
where ${\nu F_{\nu}}_{\rm tar}^{\rm obs}(E_{\rm tar}^{\rm obs})\approx4\times10^{-13}~\rm erg~s^{-1}~cm^{-2}$ is the flux of target photons, ${\nu F_{\nu}}_{\rm 20~TeV}^{\rm obs}\approx 2\times10^{-13}~\rm erg~s^{-1}~cm^{-2}$ is the flux of hard VHE spectrum as shown in Fig.~\ref{SED}, and $M_{\rm BH}\approx6.5\times10^9~M_\odot$ represents the M87's SMBH mass. It can be seen that the derived blob radius closely matches the Schwarzschild radius of M87's SMBH, suggesting that the emitting blob resides either at the jet base or at a distant location from it, while its radius remains markedly smaller than the jet's cross-section radius. A natural follow-up question is whether such a dense emitting blob can adequately account for the broadband SED. In the Thomson regime, the blob radius can be estimated through \citep{1998ApJ...509..608T}
\begin{equation}\label{TMSR}
\begin{split}
R\approx5.7\times10^{17}\rm~cm\big(\frac{\nu_h}{10^{24}\rm Hz} \big)\big(\frac{10^{14}\rm Hz}{\nu_l} \big)^2\big(\frac{3.7}{\delta_{\rm D}} \big)\\\big(\frac{L_{\rm syn}}{8.5\times10^{41}\rm erg~s^{-1}} \big)^{1/2},
\end{split}
\end{equation}
where $\nu_{l,h}$ represent the peak frequencies of low- and high-energy bumps, and $L_{\rm syn}$ is the integrated luminosity of low-energy bump. By taking $\nu_l=10^{14}\rm~Hz$, $\nu_h=10^{24}\rm~Hz$, and $L_{\rm syn}=8.5\times10^{41}\rm erg~s^{-1}$ as estimated from M87's SED, it can be found that the blob radius required to explain the VHE hard spectrum through the photomeson interactions is more than two orders of magnitude smaller than the radius needed for leptonic emission from primary electrons to account for the SED spanning optical to GeV energies. It indicates that if the VHE hard spectrum is expected to be explained by the photomeson process, then the photomeson process and the dissipation of primary electrons must occur in distinct emission regions. In this two-zone scenario, since the target photon field in the photomeson interaction region is obscured by the observed SED, the radiation spectra produced by the photomeson process remain unconstrained. Therefore, we refrain from performing SED fitting in this work. 

As discussed above, without invoking super-Eddington jet power, a compact emission region with a radius comparable to or lower than the Schwarzschild radius of the SMBH must be postulated. This requirement compels the introduction of at least a second dissipation region to account for the broadband SED of M87. Previous studies, when explaining M87's VHE radiation within the framework of the one-zone model, primarily relied on enhancing the efficiency of the photomeson process by introducing an ultra-dense soft photon fields \citep{2015EPJC...75...52S,  2022ApJ...934..158A}. However, this approach clearly overlooks the fact that no such target photon component has been observed. Alternatively, it is assumed that relativistic protons energy distribution has an exceptionally large minimum Lorentz factor to avoid the need for super-Eddington jet power \citep{2016ApJ...830...81F}. Although we have demonstrated that the photomeson interactions within the jet struggles to provide a suitable interpretation, \cite{2020ApJ...905..178K} suggested that emission of Bethe-Heitler pairs from the magnetically arrested disk (MAD) could account for the VHE spectrum of M87. They found that the emission from $pp$ interactions can be neglected in the MAD because of the low interaction efficiency.

\begin{figure}[htbp]
\includegraphics[width=0.5\textwidth]{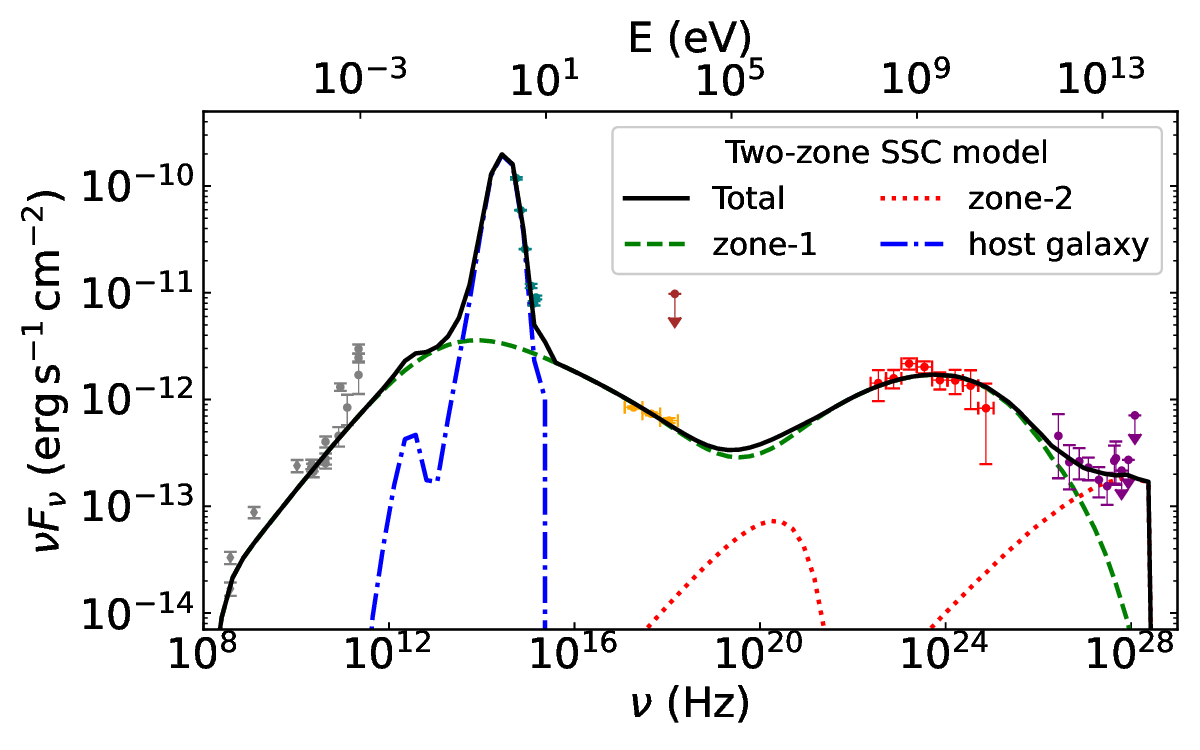}
\caption{Two-zone SSC modeling. The dashed green curve represents the leptonic emission from the first emitting region. The dotted red curve represents the leptonic emission from the second emitting region. The solid black curve is the total emission from the blob.
\label{two}}
\end{figure}

\begin{table*}
\setlength\tabcolsep{2.5pt}
\caption{Parameters of the second dissipation zone for SED Fitting with the Two-zone SSC Model.}
\centering
\begin{tabular}{ccccccc}
\hline\hline
Free parameters		&	$R$~(cm) &	$B$~(G)	&	 $L_{\rm e,inj}~(\rm erg~s^{-1})$	&	$\gamma_{\rm e,b}$	&	$p_{\rm e,1}$	 \\
\hline																							
Values  		&	$2.7\times10^{15}\rm cm$ &	$2\times10^{-4}\rm~G$	& 	$1\times10^{42}$	&	$5.0\times10^8$	&	2	\\
\hline\hline
Derived/Fixed parameters	&	$\Gamma$ &	$\theta_{\rm obs}~(\circ)$	&	$\delta_{\rm D}$		&	$\gamma_{\rm e,min}$	&	$\gamma_{\rm e,max}$&	$p_{\rm e,2}$\\
\hline	
Values&	3	&	15	&	3.7		&	1	&	$\gamma_{\rm e,b}$&	$p_{\rm e,1}$	\\
\hline
\label{twoparameters}
\end{tabular}\\
\textbf{Notes.} Values of $\Gamma$, $\theta_{\rm obs}$, $\gamma_{\rm e,min}$ and $\gamma_{\rm e,max}$ are fixed to be the same as those given in Table~\ref{ppparameters}. Since $\gamma_{\rm e,max}$ and $p_{\rm e,2}$ are poorly constrained for the second emitting zone, here we boldly assume that their values are equal to those of $\gamma_{\rm e,b}$ and $p_{\rm e,1}$ for simplicity.
\end{table*}

\subsection{Two-Zone SSC Model}
In the above subsections, we studied if the possible hard VHE spectrum can be interpreted by hadronic emissions in the framework of one-zone model. In fact, many observational evidence indicates that more than one emitting region within AGN jets can make significant contributions to the SED \citep[e.g.,][]{2023ApJ...952L..38A, 2025A&A...694A.195A, 2025ApJ...984....5L}. Therefore, the potential hard VHE spectrum in M87 may also originate from leptonic emission in a secondary emitting zone. Here we simply assume that the two emitting blobs in our two-zone model are spatially distinct. We attribute the X-ray and GeV emission to one blob and the VHE emission to the other, with no interaction between their respective emissions. In Fig.~\ref{two}, we show the fitting result of the two-zone SSC model. The parameters used in the first emitting blob are the same as those given in Table~\ref{ppparameters}, and the parameters used in the second emitting blob are given in Table~\ref{twoparameters}. With the introduction of a larger $\gamma_{\rm e, b}$ than that of the first dissipation zone (to overcome the KN effect), the SSC emission from the second dissipation zone can account for the possible hard VHE spectral feature detected by LHAASO.

\section{Discussion and conclusion}\label{DC}
The detection of a possible hard VHE spectrum at $\sim$20 TeV in M87 by LHAASO challenges the canonical one-zone SSC model and necessitates alternative radiation mechanisms. In this work, we explore both hadronic and two-zone SSC models to interpret the broadband SED, particularly focusing on the origin of the hard VHE component. We find that the $\pi^0$ decay gamma rays in $pp$ interactions successfully reproduce the hard VHE spectrum in the framework of one-zone model. This models operate within sub-Eddington jet power constraints, ensuring physical plausibility. {With analytical method, we find that the p-syn radiation cannot interpret the hard VHE spectrum.} The photomeson model also faces significant spatial constraints. To explain the VHE hardening via $\pi^0$ decay gamma rays, the emitting region must be compact ($R\sim R_S$), yet such a region cannot simultaneously account for the broadband SED from optical to GeV energies. This necessitates at least two distinct emission zones: one for the primary leptonic component and another for the photomeson-driven VHE emission. However, the absence of observed target photon fields in the required energy range ($\sim$21 keV) further complicates this scenario, making it less favored without additional ad hoc assumptions. A two-zone SSC model, where a secondary emission region with a higher break electron Lorentz factor contributes to the VHE spectrum, also provides a viable explanation. This setup mitigates the KN suppression at high energies but requires extreme electron acceleration efficiency. 

For all the phenomenological models tested in this work, the strength of the magnetic field exhibits significant variations, indicating substantial differences in the location of the emitting blob.  For both the one-zone SSC+$pp$ model and the two-zone SSC model, the broadband SED except the hard VHE spectrum detected by LHAASO originates from leptonic emission contributed by a single electron population. As shown in Eq.~(\ref{TMSR}), the relatively large ratio of the peak frequency of the high-energy bump to that of the low-energy bump necessitates a substantially large blob radius. Given that the fluxes of the high-energy and low-energy peaks are comparable, the derived magnetic field strength is also relatively weak, on the order of $\rm mG$, which is consistent with values obtained in previous studies applying one-zone leptonic models \citep{2009ApJ...707...55A, 2020MNRAS.492.5354M, 2021ApJ...911L..11E}. It should be noted that mG magnetic fields conflict with radio observations, which indicate that the magnetic field strength at deprojected distances $\lesssim100~R_S$ is approximately $\gtrsim1\rm~G$ \citep[e.g.,][]{2015ApJ...803...30K, 2016ApJ...817..131H, 2021ApJ...911L..11E}. This indicates that the radio core and the VHE emission likely originate from distinct emission regions. If accounting for sub-TeV emission with p-syn model, the necessity for a strong magnetic field arises from p-syn radiation dominating the high-energy bump. As illustrated in the lower panel of Fig.~\ref{psyn}, the magnetic field strength is constrained to $\sim 2\rm~G$, which is generally consistent with previous studies \citep{2004A&A...419...89R, 2020MNRAS.492.5354M}. The required magnetic field strength varies significantly across different models, implying considerable discrepancy in the inferred location of the emitting blob responsible for producing M87's hard VHE spectrum. Based on observations of KVN, VERA Array, and the Very Long Baseline Array, \cite{2023A&A...673A.159R} proposed a relation between magnetic field strength and the deprojected distance $z$ from the SMBH, which is $B=(0.3-1.0\rm~G)(\it{z}/\rm450\it{R}_S)^{\rm-0.73}$. Therefore, if the VHE radiation originates from the one-zone SSC+$pp$ model or the two-zone SSC model with a $\rm mG$ magnetic field, the emission region would be located at approximately $10^6~R_S$, proximal to the HST-1 knot ($\sim10^5~R_S$). If p-syn radiation dominates the VHE radiation, the emitting blob would reside at several hundred $R_S$, corresponding to sub-pc scales. Moreover, the p-syn scenario predicts a higher X-ray polarization degree compared to the SSC emission \citep{2013ApJ...774...18Z, 2024ApJ...967...93Z}. Current \citep[e.g., IXPE;][]{2016SPIE.9905E..17W} or future \citep[e.g., eXTP;][]{2025arXiv250608101Z} X-ray polarimetry observations could test this prediction and distinguish between these scenarios. 

\section*{Acknowledgements}
We thank the anonymous referee for insightful comments and constructive suggestions. 
This paper employs a list of Chandra datasets, obtained by the Chandra X-ray Observatory, contained in the Chandra Data Collection ~\dataset[DOI: 10.25574/cdc.430]{https://doi.org/10.25574/cdc.430}.
This work is supported by the National Key R\&D Program of China (2023YFB4503300), the National Natural Science Foundation of China (NSFC) under the grants No. 12203043, No. 12203024, and No. 12473020, the Yunnan Province Youth Top Talent Project (Grant No. YNWR-QNBJ-2020-116), and the CAS ``Light of West China" Program.

\appendix
\section{Fitting results with another value of $\Gamma$}\label{APP}
\begin{figure*}[htbp]
\subfigure{
\includegraphics[width=0.5\textwidth]{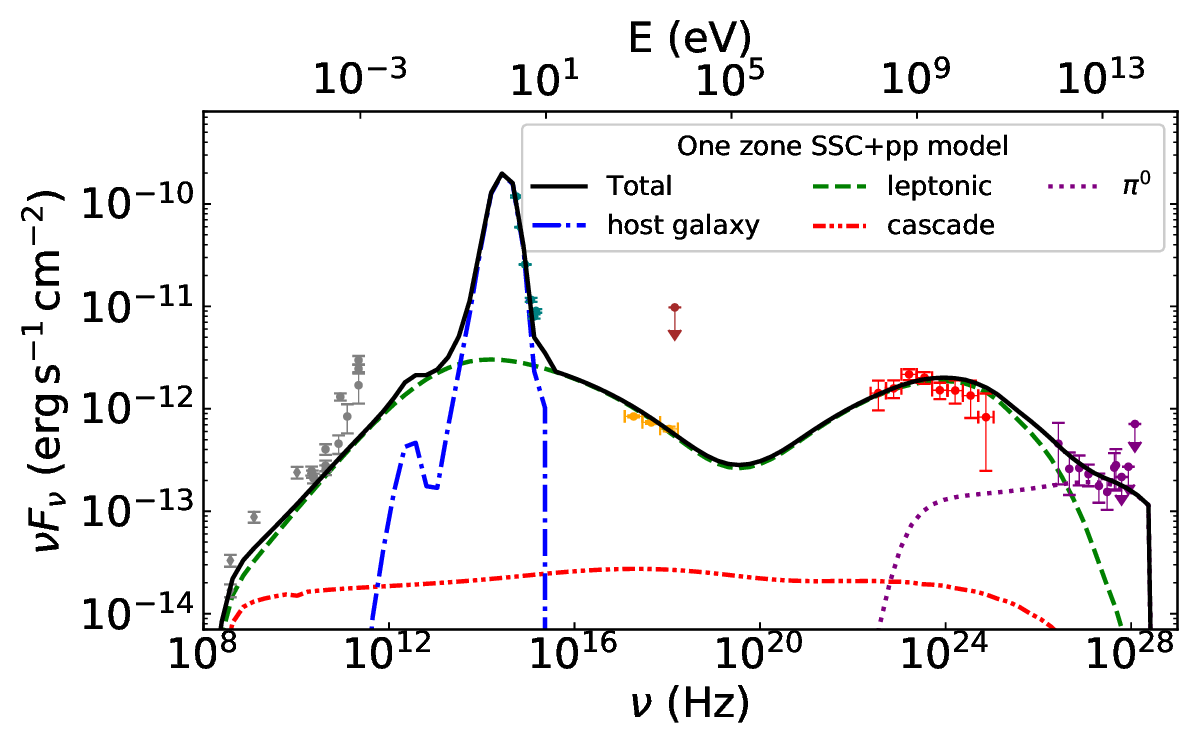}
}\hspace{-5mm}
\quad
\subfigure{
\includegraphics[width=0.5\textwidth]{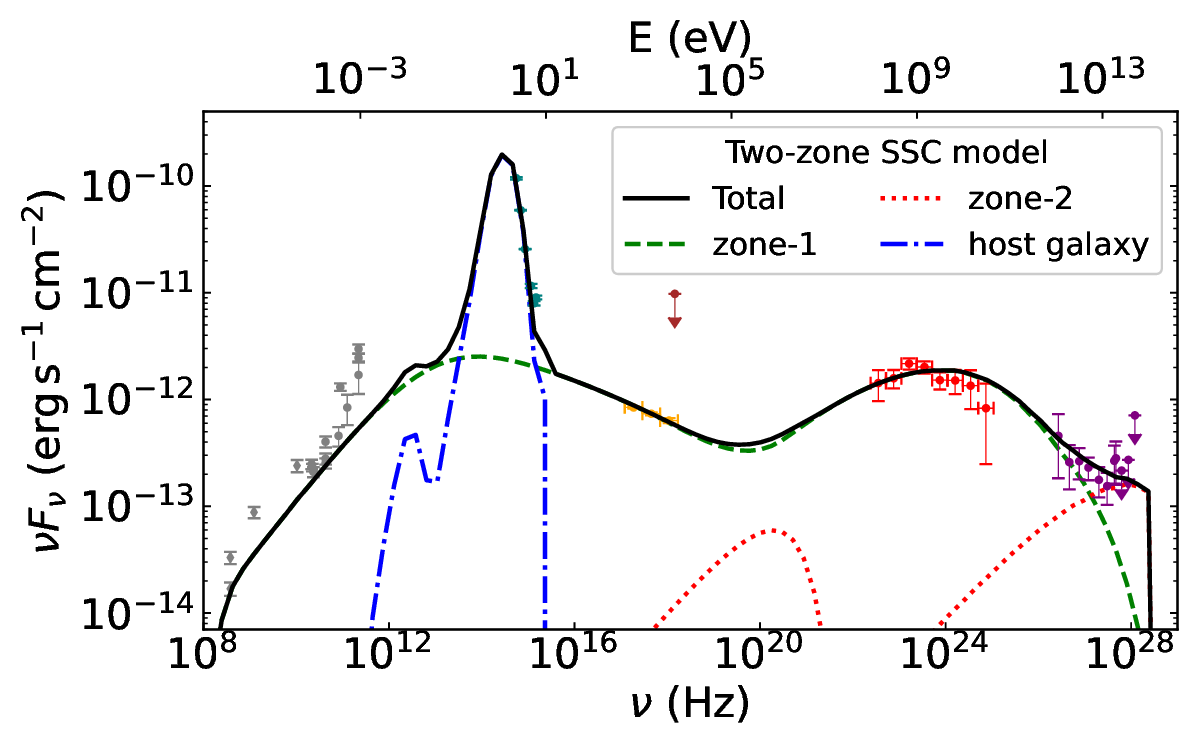}
}
\caption{Fitting results of one-zone SSC+$pp$ model (left panel) and two-zone SSC model (right panel) with $\Gamma=10$. All the data points and line styles have the same meaning as in Figures~\ref{pp} and \ref{two}.
\label{sumfig}}
\end{figure*}

In our default SED fitting shown in Figs.~\ref{pp} and \ref{two}, we fix $\Gamma=3$, which is usually applied in previous modeling studies and suggested by $\sim$pc scale observations. Even though, some studies \citep[e.g.,][]{2016A&A...595A..54M, 2019ApJ...887..147P} reveal that the jet of M87 continues to accelerate to larger scales, with measured $\Gamma \lesssim10$. Therefore, it is useful to test if the models applied in this work can also fit the averaged multiwavelength SED of M87 well with a larger $\Gamma$. Here, we show the fitting results of one-zone SSC+$pp$ model and two-zone SSC model with $\Gamma=10$ in Fig.~\ref{sumfig}. Compared to previous modelings, adjustments were made to several parameters. For the one-zone SSC+$pp$ modeling, when setting $\Gamma=10$, the upper limit of $R$ given by Eq.~(\ref{RS}) is $7.3\times10^{18}\rm~cm$. It can be seen that a substantial parameter space still remains for the $pp$ interactions to interpret the hard VHE spectrum at $\sim20\rm~TeV$. In the modeling, we set $B=1.5\times10^{-3}\rm~G$, $L_{\rm e,inj}=1.5\times10^{43}\rm~erg~s^{-1}$, $p_{\rm e,2}=3.28$, $\xi=0.1$, $\gamma_{\rm p,max}=1\times10^6$, with all other parameters consistent with Table~\ref{ppparameters}. For two-zone SSC modeling, we set $L_{\rm e,inj}=2.1\times10^{43}\rm~erg~s^{-1}$, with all other parameters consistent with Table~\ref{twoparameters}, because when setting $\Gamma = 10$, the corresponding $\delta_{\rm D} \approx 2.6$ differs slightly from the previous value and increasing $L_{\rm e,inj}$ suffices to compensate for the flux reduction caused by a decrease in $\delta_{\rm D}$.

\vspace{5mm}
\facilities{Chandra, \textsl{Swift}-XRT, \textsl{Swift}-UVOT}

\software{astropy \citep{2013A&A...558A..33A,2018AJ....156..123A, 2022ApJ...935..167A},  
          naima \citep{naima}
          }
          
\bibliography{ms.bib}{}
\bibliographystyle{aasjournal}
\end{document}